\documentclass[usenatbib]{mnras}

\usepackage{amsmath}
\usepackage{siunitx}
\usepackage{xfrac}
\usepackage{natbib}
\usepackage{xcolor}
\usepackage{graphicx}
\usepackage{booktabs}
\usepackage{csquotes}
\usepackage{hyperref}            
            
\hypersetup{pdfpagemode = {UseNone},
            pdftitle = {Parameter study},
            pdfcreator = {\LaTeX},
            pdfproducer = {pdfeTeX-0.\the\pdftexversion\pdftexrevision},
            pdfauthor = {Anna Penzlin},
            pdfsubject = {},
            pdfview = {FitH},
            pdfstartview = {FitH},
            colorlinks = {true},
            linkcolor = [rgb]{0,0.35,0.7},
            citecolor = [rgb]{0,0.35,0.7},
            filecolor = [rgb]{0.61,0,0},
            urlcolor = [rgb]{0.61,0,0},
           }
           
\usepackage{bm}
\renewcommand{\vec}[1]{\bm{\mathrm{#1}}}

\colorlet{darkgreen}{green!60!black}

\newcommand{\ab}{a_\mathrm{bin}}
\newcommand{\eb}{e_\mathrm{bin}}
\newcommand{\Tb}{T_\mathrm{bin}}

\newcommand{\beq}{\begin{equation}}
\newcommand{\eeq}{\end{equation}}

\title[I - Structure of the inner circumbinary cavity]{ Viscous circumbinary protoplanetary discs - I.
Structure of the inner cavity}

\author[Penzlin et al.]{
Anna~B.T. Penzlin,$^{1}$\thanks{E-mail: a.penzlin@imperial.ac.uk}
Richard~A. Booth$^{2}$,
Richard~P. Nelson$^{3}$,		
Christoph~M.~Sch\"afer$^{4}$,
Wilhelm Kley\thanks{passed away}
\\
$^{1}$Astrophysics Group, Department of Physics, Imperial College London, Prince Consort Rd, London, SW7 2AZ, UK \\
$^{2}$School of Physics and Astronomy, University of Leeds, Leeds LS2 9JT, UK\\
$^{3}$Astronomy Unit, Department of Physics and Astronomy, Queen Mary University of London, London E1 4NS, UK\\
$^{4}$Institut f\"ur Astronomie und Astrophysik, Universität T\"ubingen,
Auf der Morgenstelle 10, 72076 T\"ubingen, Germany
}

\date{Accepted XXX. Received YYY; in original form ZZZ}

\pubyear{2024}

\begin{document}
\label{firstpage}
\pagerange{\pageref{firstpage}--\pageref{lastpage}}
\maketitle

\begin{abstract}
{Many of the most intriguing features, including spirals and cavities, in the current disc observations are found in binary systems like GG Tau, HD 142527 or HD 100453. 
Such features are evidence of the dynamic interaction between binary stars and the viscous disc. 
Understanding these dynamic interactions and how they result in the structure and growth of asymmetric circumbinary discs is a difficult problem, for which there is no complete analytical solution, that predicts the shape of the observed disc accurately.
We use numeric simulation to evolve circumbinary discs with varying disc viscosities and investigate the size and shape of the inner cavities in such protoplanetary discs.
We have simulated over 140 locally isothermal 2D grid-based disc models for $\geq3\times10^4$ binary orbits each and mapped out the parameter space relevant for protoplanetary discs.
With this, it becomes possible to create parametrised profiles for individual discs to compare to observation and find limits to their binary eccentricity or internal viscosity from the simulation data.
In the long-term simulations larger cavity sizes than previously considered are possible within the parameter space ($\leq $ 6 binary separations).
As an example, we find that the eccentricity of the disc around HD 142527 suggests the impact of the binary dynamics on the disc. However, even considering the larger cavity sizes, the large size of the cavity in HD 142527 remains unexplained by the simulations considering the most recent orbital constraints.
}\end{abstract}


\section{Introduction}

Half of all stars are found in binary or multiple configurations \citep{2010solar-binaries,2022PPVII} and often also form together surrounded by one common circumbinary disc (CBD). Famous examples of resolved circumbinary discs include GG Tau \citep[e.g.][]{1999Guilloteau} and HD 142527 \citep{2006Fukagawa,2011Verhoeff}.
Observations of such discs reveal numerous special features and structures like cavities, spirals, streams and shadows, eccentric disc edges \citep[e.g.][]{2020Keppler,2021HD142527} and dust over-densities \citep{2015Casassus}.
Furthermore, about a dozen circumbinary planets have been detected that arise from circumbinary protoplanetary discs (PPDs). Most planets are located near the instability limit \citep{2018Quarles} described by \cite{1999Holman}. 

The features of the observations can help to limit and predict properties of the disc that are not directly observable. For example, the best fit observational constraints for GG Tau in cavity size and stellar orbits can be achieved by simply choosing a low viscosity and evolving the disc simulations for long enough, as shown in \cite{2020Keppler}.
As disc properties change for different length scales and times in their evolution, we need a large parameter space to compare observations to and find the best fitting parameters for each system, which can help us also to constrain the disc and orbital parameters better from observations of circumbinary discs.
 
To understand the dynamics that are responsible for these features, we need hydrodynamic models of viscous, pressure-supported discs 
affected by the interaction of a temporally changing stellar gravitational potential. Results of the theoretical work by \cite{2020Munoz_theo} of binary-disc interaction can explain the excitation of eccentricity in the inner cavity. 
 However the exact disc shape and interaction with the binary is sensitive to disc conditions such as temperature and viscosity.

 While previous studies of protoplanetary, circumbinary discs \citep{2017Mutter, Calcino_2019, 2020Hirsh, 2020Ragusa, 2021Pierens, 2022Penzlin} agree on the development of large, eccentric cavities, how disc and binary parameters contribute to the final cavity shape
 is not yet fully understood.
Our goal is to create a general picture of how the circumbinary disc structure depends on disc viscosity, scale height and binary eccentricity using numerical simulations.
This can provide a translation map for observed discs to extract more disc or binary parameters from the observable gas properties in observed young binary systems and eventually provide a better understanding of the evolution of the whole parameter set during the disc evolution.

\section{Models}\label{sec:model}
	
	In this work we use finite-volume simulations to investigate the binary-disc interaction. This method allows
	parameterisation of the viscosity through constant values of aspect ratio and Shakura-Sunyaev parameter $\alpha$ \citep{alpha1973}
 	throughout the domain. 
	
	\subsection{Setup}
	
We use the PLUTO code \citep{2007Pluto} with the GPU-capable version by \cite{2017Thun} to simulate the 2D hydrodynamical gas evolution around the binaries. We use a locally isothermal equation of state and a non-flaring disc to further simplify and parameterize the disc. 

The kinematic viscosity in our models can be written as
	\begin{equation}
	\label{eq:nu}
	\nu = \alpha c_s H \,.
	\end{equation}
The parameter $\alpha < 1$ describes the turbulent viscosity, $c_s$ the adiabatic sound speed of the gas, and $H$, the vertical pressure scale height. 
For simple $\alpha$ models, observations \citep{2018Dsharp6}
suggest an $\alpha$-value between $10^{-2}$ and $10^{-4}$. We will typically take $\alpha$ to be $10^{-3}$, if not stated otherwise.
	
The setup follows \cite{2017Thun}.
The resolution of our cylindrical grid is $684\times 574$. Our radial domain extends from $1 a_\mathrm{bin}$ to $40 a_\mathrm{bin}$ with logarithmic spacing.
\cite{2017Thun} demonstrated that this resolution resolves the disc structure. 
Tests of $4\times$ resolution with low viscosity using $\alpha=10^{-4}$ showed no significant changes.
We exclude the stars from the domain to enable us to simulate the system to a steady converged disc structure in reasonable computation times.
We tested a smaller inner boundary ($0.75~a_\mathrm{bin}$) as in \cite{2022Penzlin}, and found that the the structure of cavity and inner disc are robust. 
	
The inner boundary is a diode boundary, as the boundary is well within the opened cavity and compared to other simulations by \cite{2021Tiede} accretion of $>90\%$ of disc material entering this region is expected. The outer boundary damps to the initial profile in the last 10\% of the simulation domain leading to a fixed outer profile.
	The floor density is $10^{-6}$ times the initial density.
	
	The initial surface density, $\Sigma(t=0)$, is 
    given by 
    \begin{equation}\label{eq:init_sigma}
	\Sigma(t=0) = f_\mathrm{gap} \Sigma_\mathrm{ref} \left(
	\frac{R}{a_\mathrm{bin}} \right)^{-1.5} \,,
	\end{equation}
     where $f_\mathrm{gap}$ is a factor used to reduce the surface density in the region around the binary, described in \cite{2017Thun}:
	\begin{equation}
	f_\mathrm{gap} = \left[1+\exp{\left(-\frac{R-R_\mathrm{gap}}
		{\Delta R}\right)} \right]^{-1} \,,
	\end{equation}
	with $R_\mathrm{gap} = 2.5\,a_\mathrm{bin}$ and $\Delta R =
	0.1\,R_\mathrm{gap}$. The reference density $\Sigma_\mathrm{ref}=1.67\times 10^{-4}~M_\mathrm{bin} \ab^{-2} $ is to produce a disc mass of $M_\mathrm{disc}=0.01 ~M_\mathrm{bin}$ inside the computational domain. In all models, the effects of the disc on the binary are not included; disc self-gravity is also not included.
	
 We ran our simulations for $30\,000~\Tb$ or until the cavity size remained steady apart from periodic variations caused by the precession of the disc, whichever was longer.
 
Any features seen before the disc reaches its full cavity size can be caused by the overly symmetric and smooth, initial gas distribution. 
Other works \citep[e.g.][]{Calcino_2019} report gas overdensities that orbit near the inner edge independently from the precession of the overall disc, called "lumps".
Early in the simulations after $\sim 500 - 2000~\Tb$, lumps form from the material initialized in the unstable region, that gets pushed outwards into the disc. 
In our models, these lumps mainly appear for higher viscosity and dissolve within a few thousand orbits.
Some simulations are not able to overcome the symmetric initial condition for up to $\sim 5000~\Tb$ and show meta-stable circular inner cavities especially if the binary masses are similar \citep[see also []{2020Keppler}. However, tiny perturbations are enough to overcome circular inner cavities and initiate cavity growth and excitation. 

Once the simulations have reached steady-state, every disc shows a large, eccentric, precessing cavity of more than $3~\ab$.

	\begin{table}[ht] 
		\caption{Parameters for all hydrodynamic simulations. $q=M_2/M_1$ is the mass ratio, $\eb$ the eccentricity of the binary orbit, $\alpha$ the viscous parameter and $h$ the disc scale height.
		}
		\centering 
		\begin{tabular}{l c c c c} 
			\hline\hline 
			\rule{0pt}{5mm}
			Models & q & $\eb$ & $\alpha$ & $h$ \\ [1ex] 
			\hline
			Fiducial & 0.26 & 0.2 & $10^{-3}$ & 0.05 \\
		    $\alpha$ variation& 0.26 & 0 - 0.4 & $10^{-4}-10^{-2}$ & 0.05\\
			$h$ variation & 0.26 & 0 - 0.4 & $10^{-3}$ & 0.03 - 0.1\\
            Thick discs & 0.26 & 0 - 0.4 & $10^{-4}, 10^{-3}, 10^{-2}$ & 0.1\\
			\hline 
		\end{tabular}
		\label{table: hydro parameters} 
	\end{table}
 
	We varied the aspect ratio, viscous $\alpha$ parameter and binary mass ratio for binary eccentricities between 0 and 0.4. As fiducial values we used  $h =0.05$, $\alpha = 10^{-3}$, and a mass ratio of $q=M_2/M_1=0.26$ (see Tab.~\ref{table: hydro parameters}). These values were chosen based on another study of the formation of Kepler-38b. The mass ratio is crucial for the way that mass is accreted between the binary components \citep{2014Farris, 2020Duffell, 2023siwekI}. 
	However, the cavity and inner disc structure are less sensitive to the mass ratio, \cite{2018Thun} found comparable cavity shapes for $q = [0.29, 0.60, 0.90]$ (systems like Kepler-16, -34 and -413).
 
	\subsection{Characterization of the disc shape}
	
To parameterize the structure of the cavity, we define the cavity edge based on the location where the density is $10\%$ of the maximal density at the same azimuth. 
As in \cite{2017Thun}, the argument of apocentre of the inner disc is identified as the azimuthal position of the maximum density beyond the unstable region \citep{1999Holman}. 
The positions of the pericentre and apocentre of the cavity are then identified as the position of 10\% of the maximum density at corresponding angles. These positions, together with the centre of mass as the focal point, define the shape of the cavity edge. 
In the following, the cavity semi-major axis $a_\mathrm{cav}$ and eccentricity $e_\mathrm{cav}$ are calculated from this ellipse as a simple parametrized measure of the cavity shape.
 
	We also define the eccentricity vector $\vec{e}$ throughout the disc following \cite{2017Miranda}.
	The eccentricity vector $\vec{e}$ indicates the orbit alignment of the disc through the extent of the disc.  
	It is derived using the the local velocity $\vec{v}$ and position $\vec{r}$ in the following way:
	\begin{equation}
	\vec{e} = \left(e_x \atop e_y  \right)= \frac{1}{GM}(\vec{v}^2\vec{r}-(\vec{v}\cdot\vec{r})\vec{v}) - \frac{\vec{r}}{|r|}. \label{eq:ecc}
	\end{equation}
	
	We use the mass-weight absolute eccentricity $\langle\vec{e}\rangle$ described in \cite{2017Miranda} for the eccentricity of the gas locally.
	The local radial argument of pericentre $\omega$ can thereby be calculated by
	\begin{equation}
	\omega = \arctan2(\langle e_y\rangle/\langle e_x \rangle)
	\end{equation}
 
	Previous work \citep{2017Miranda} shows that the elliptic orbits introduced by the disc extend deep into the disc ($\sim 10~a_\mathrm{bin}$) with a rigid precession of all orbits around the centre of mass. 

 To compare to observations more easily, we also describe the orbit of the peak density in the apocentre of the disc with $a_\mathrm{peak}$ and $e_\mathrm{peak}$ later in the paper. This should be traced by scattered light observations. For continuum observation, dust is trapped by the gas density maxima and gets stirred up beyond this position \citep{2020Pierens, 2021Pierens}. Therefore the observed signals tend to be beyond scattered light signals that trace the gas dynamics more closely.
	
	\section{The general disc behavior}\label{sec:result}

\begin{figure}
		\resizebox{\hsize}{!}{\includegraphics{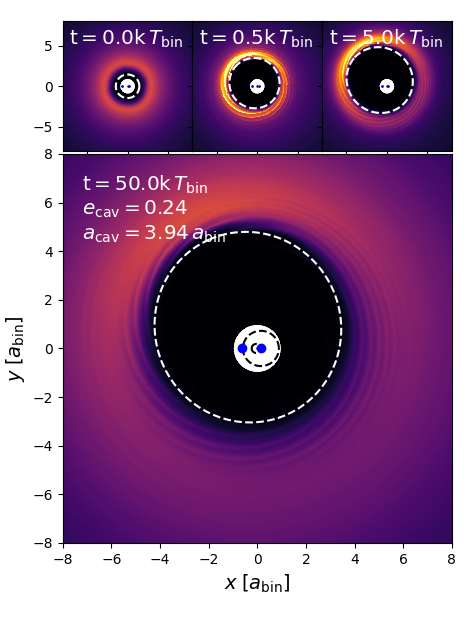}}
		\caption{2D gas density plot of the "standard disc"($\alpha=10^{-3},\,H/R=0.05,\, e_\mathrm{bin}=0.2$) simulated for 0, 500, 5000 and $50\,000 \, T_\mathrm{bin}$. The white area marks the inner boundary. The binary position is denoted by the blue dots. The elements of the elliptic cavity(dashed white line) can be found in the top left corner.}
		\label{fig:2d}
	\end{figure}
 
	\begin{figure}
		\resizebox{\hsize}{!}{\includegraphics{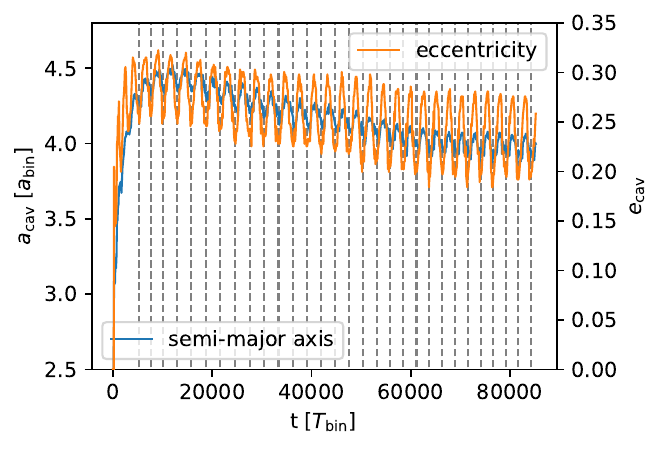}}
		\caption{Cavity evolution of the "standard disc"($\alpha=10^{-3},\,H/R=0.05,\, e_\mathrm{bin}=0.2$) simulated over $85\,000 \, T_\mathrm{bin}$. The blue line is the semi-major axis of the cavity and the orange line marks the cavity eccentricity. The grey dashed lines indicate when the pericentre of binary and disc is anti-aligned($\Delta\omega=\pi$).}
		\label{fig:cav_evo}
	\end{figure}
	
	When simulated to a (quasi-)steady-state, circumbinary discs show a large, eccentric, inner cavity with a semi-major axis $\geq 2.5 \, a_\mathrm{bin}$ that depends on the disc parameters. For the fiducial model, cavity semi-axis reaches $4 \, a_\mathrm{bin}$ as shown in Fig.~\ref{fig:2d}.
 
 \subsection{Converging the disc simulation}

The snapshots in Fig.~\ref{fig:2d} illustrate that the disc structure changes over long time scales with slowing rate to eventually converge to a final disc configuration.
Once the disc has reached a convergent state after $\geq 50\,000\,T_\mathrm{bin}$,
the inner disc shows spiral features driven by the binary motion, with the maximum density at the apoastron due to the eccentric, near-Keplerian motion of the gas. Otherwise, we find no over-density or asymmetries in the gas density for protoplanetary CBDs (implying a low viscosity environment). We do see overdensities moving around the cavity edge, similar to the lumps reported in other works \citep[e.g.,][]{Calcino_2019}, after shorter simulation times, e.g., at stages comparable to the $500~\Tb$ panel in Fig.~\ref{fig:2d} when the binary still clears its inner cavity and displaces a significant amount of gas. But only in for thicker discs ($h > 0.06$) do these remain noticeable for after $5000\Tb$. After $10\,000~\Tb$ we do not notice any more lumps in our simulation set. The evolution of the cavity in Fig.~\ref{fig:cav_evo} shows that the disc needs $\approx 10^5~\Tb$ to fully overcome the symmetric initial condition. The oscillations in eccentricity and semi-major axis present at the end of the simulation are related to the disc precession and are discussed further in the next section.
Hence, there is not one steady state of the final disc but rather an oscillation around one mean value.

	\begin{figure*}
		\resizebox{\hsize}{!}{\includegraphics{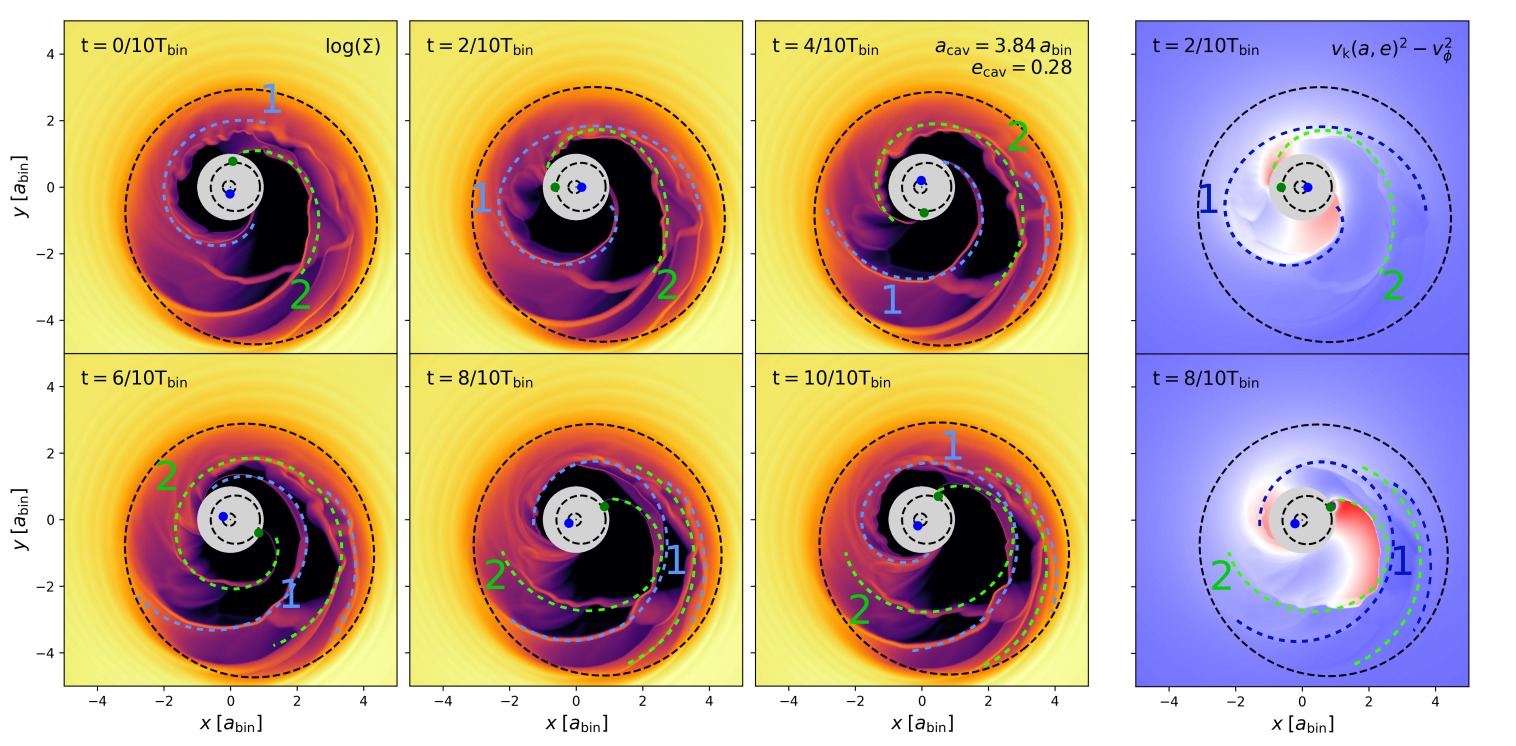}}
		\caption{2D logarithmic gas density map (left 6 panels) and square of the radial eccentric velocity component (right 2 panels)  of the "standard disc"($\alpha=10^{-3},\,H/R=0.05,\, e_\mathrm{bin}=0.2$) over one orbit $T_\mathrm{bin}$. The grey area marks the inner boundary. The binary positions are denoted by the green(secondary) and blue(primary) dots and the wakes caused by either with dashed coloured lines. In the velocity panels (right), blue indicates velocities that are on a bound Keplerian orbit and red signals super-Keplerian outflow velocities, even when accounting for eccentric Kepler orbits for the gas. The dashed blue/green spiral lines qualitatively trace the wake to the star causing the motion.}
		\label{fig:orbit_evo}
	\end{figure*}
 
	\subsection{Example of short-term and precession dependent evolution}
	\label{sec:short_term}
	
Figure \ref{fig:orbit_evo} shows the behaviour of the gas during one binary orbit. 
The primary (blue) and the secondary (green) both launch spiral wakes due to the motion of the potential. 
When one of the stars approaches the pericentre of the cavity it pulls gas towards the binary. This leads to a periodic accretion pattern with the strongest peak at the primary passage through the cavity pericentre.
Such a material inflow can be seen near the secondary position at $t=2/10~\Tb$. 
As the secondary has a wider orbit but the same angular velocity as the primary, the material inside the wake following the secondary rotation collides with the primary wake, reducing the overall outwards motion of the material pushed by both spiral wakes. 
This outwards motion of the wakes can be seen in the undisturbed portion of the primary wake that expands outwards towards the disc's apocentre and lights up with a red, super-Keplerian, radial velocity component inside its local orbit in the right panels.
At $t=6/10~\Tb$ the primary reaches a position near the apocenter and material streams out of the domain. Meanwhile the secondary wake now again catches up with the primary wake near the apocenter of the disc and collides at $t=8/10~\Tb$, recreating the initial multi-spiral pattern in the disc. The high velocity difference between primary and secondary wake becomes fully clear when comparing the wakes' super-Keplerian, radial velocities.
When the primary passes its pericentre at $t=2/10~\Tb$ it causes a slower outwards motion than the secondary at its apocentre passage at $t=8/10~\Tb$.
The material and therefore momentum of both wakes depends on the overall accretion which is determined by the discs viscosity \citep{2022Penzlin}, thereby the strength of turbulence but also the thermal profile and conditions inside and outside of the cavity become relevant for the final disc shape \citep{2022Sudarshan}.

At the time of the snapshots, the binary stars mainly accrete from the top left where the pericentre of the disc is currently located.
This illustrates how the angle between the argument of pericentre of the binary and cavity $\Delta\omega$ leads to an asymmetric configuration.
This relative angle changes over time due to the slow precession of the disc. This means that the closest point between binary and disc varies and with the distance, the maximum gravitational forces pulling and exciting disc material during its orbit varies. If $\Delta\omega=0$, the binary and the disc are aligned and the closest distance is between the pericentre of the disc and the pericentre of the secondary. If $\Delta\omega=\pi$, the binary and the disc are anti-aligned and the closest distance is between the pericentre of the disc and the apocentre of the secondary. Thereby the closest distance changes by $a_\mathrm{cav}(1-e_\mathrm{cav}) - a_2(1-\eb)$ to $a_\mathrm{cav}(1-e_\mathrm{cav}) - a_2(1+\eb)$ during one disc precession, which for our highest binary eccentricity of 0.4 and a circular cavity at 3.5 would translate to $\sim \pm 30\%$ variation around the mean distance between disc edge and secondary star. Thereby, the precession becomes relevant.
The disc-binary alignment becomes most relevant for the gravitational torque between disc and binary, which will be highlighted in the second article linked to this study. 

The distance between the secondary pericentre and the cavity edge affects the shape and size of the cavity as well. This is reflected in the oscillating pattern of the cavity evolution in Fig.~\ref{fig:cav_evo}, for which the frequency matches the rigid precession of the inner disc. 


\section{Parameter investigation}\label{sec:result_II}

The shape of circumbinary discs depends on the binary and disc parameters. First, the two most relevant binary parameters in scale-free models are the binary mass ratio and the binary eccentricity. The disc shape is nearly insensitive to mass ratio, $q$, for $q = 0.2 - 1$ \citep{2018Thun,2020Hirsh}. At $q < 0.1$ the cavity size becomes noticeably smaller as the secondary becomes more planet-like. \cite{2016q-transition} found a transition to a planet-like case at $q = 0.04$. In this study we focus on the regime that allows the full growth of the cavity. 
The binary eccentricity has a more significant, and non-monotonic, effect on the size and shape of the inner disc \citep{2017Thun}.

	\subsection{Binary eccentricity}

\begin{figure}
		\centering
		\resizebox{\hsize}{!}{\includegraphics{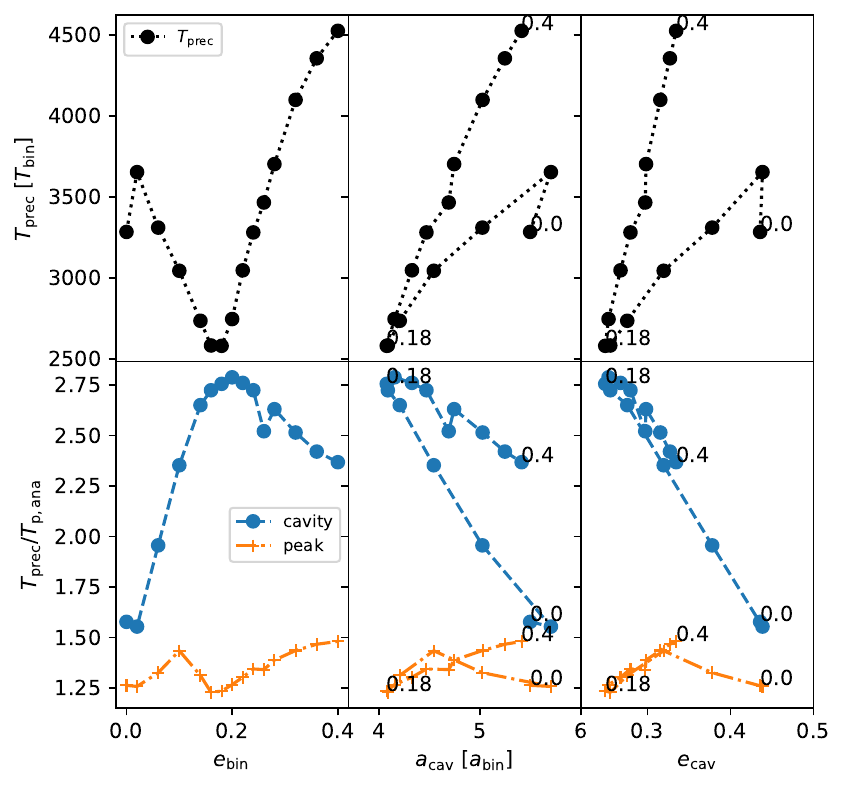}}
		\caption{The change in precession time (black) with varying binary eccentricity (left) and cavity size (mid) and eccentricity (right) compared to the precession normalized by the theoretical orbit precession at the cavity position (blue) as described in \eqref{scale}. The small numbers (mid/right) indicate binary eccentricity. Viscous $\alpha = 10^{-3}$ and the disc scale height is $h=0.05$.}
		\label{fig:mass_period}
	\end{figure}

In addition to the size and shape of the cavity, also the precession time of the disc changes non-monotonically with the binary eccentricity.
The upper row of Fig.~\ref{fig:mass_period} shows how the precession time and shape of the cavity edge depend on the binary eccentricity for viscous $\alpha = 10^{-3}$. 
The eccentricity of the cavity always exceeds 0.2 and the cavity size is $a_\mathrm{cav}>4~\ab$.
The cavities are all much larger and more eccentric than the classical results found in the estimation and simulations by \cite{1994Arty}, which compared early phase cavity clear in hydrodynamic simulations to the position of the Lindblad resonances that are responsible for gap opening of massive planets.
These results by \cite{1994Arty} are rather a lower bound that marks the unstable region around the binary. 
The eccentric modes in the disc are long-lived but also slow to excite \citep{2020Munoz}.
\cite{2020Hirsh} reaches comparable disc structures using the smooth particle hydrodynamics for the protoplanetary disc environment.

As often discussed \citep[e.g.][]{2017Miranda, 2022Dittmann}, the inner disc precesses rigidly, such that the cavity's argument of pericentre can be used to describe the inner disc as a whole.
The disc precession takes between $2500~\Tb<T_\mathrm{prec}<4600~\Tb$ and leads to the oscillation of the disc shape on this time scale as shown in vertical lines in Fig.~\ref{fig:cav_evo}. 
We compare the precession period of the disc $T_\mathrm{prec}$ with the analytic precession rate of a mass-less test particle orbiting the binary \citep{2004Moriwaki}:

	\begin{equation}
	T_\mathrm{p,ana} = \frac{4}{3} \frac{(q+1)^2}{q} a_\mathrm{obj}^{3.5} \frac{(1-e_\mathrm{obj}^2)^2}{1+1.5 e_\mathrm{bin}^2} T_\mathrm{bin}. \label{scale}
	\end{equation}
    The precession period of the cavity is however expected to exceed this estimate substantially because most of the disc mass is further from the binary than the cavity inner edge. 
	
The precession time is about $\sim2.5$ times longer than the precession of a massless particle at the cavity edge. Between a binary eccentricity of 0 and 0.16 this factor increases linearly and varies little for larger eccentricities. This change of behaviour and bimodal precession periods at $\eb \sim 0.18$ can be observed for the viscous $\alpha=10^{-3}$ as before for higher viscosities \citep[e.g. $\alpha=10^{-2}$][]{2017Thun}. 
Comparing the precession period to the 3-body precession time evaluated at the peak of the surface density provides a closer match.
Because the cavity edge has a low mass, while the peak orbit contains a more significant amount of the rigidly precessing inner disc mass, this is an expected effect.
The disc precession is still slower than the theoretical 3-body orbit at the density peak
by a factor of 1.25. This indicates that the theoretical orbit which matches the precession of the inner disc lies further out {due to the further extend of the rigid precession}. 
However, in contrast to matching the precession to the position of the cavity, the match of theoretical precession to disc precession at the peak density location depends much less on the binary eccentricity. This can be related to the narrow width between peak density and cavity edge for low eccentricity binaries. Therefore, the precession time will also shift, if the properties inside of the disc lead to a changing extent of the material trapped in precession.

	\begin{figure*}
		\centering
		\resizebox{\hsize}{!}{\includegraphics{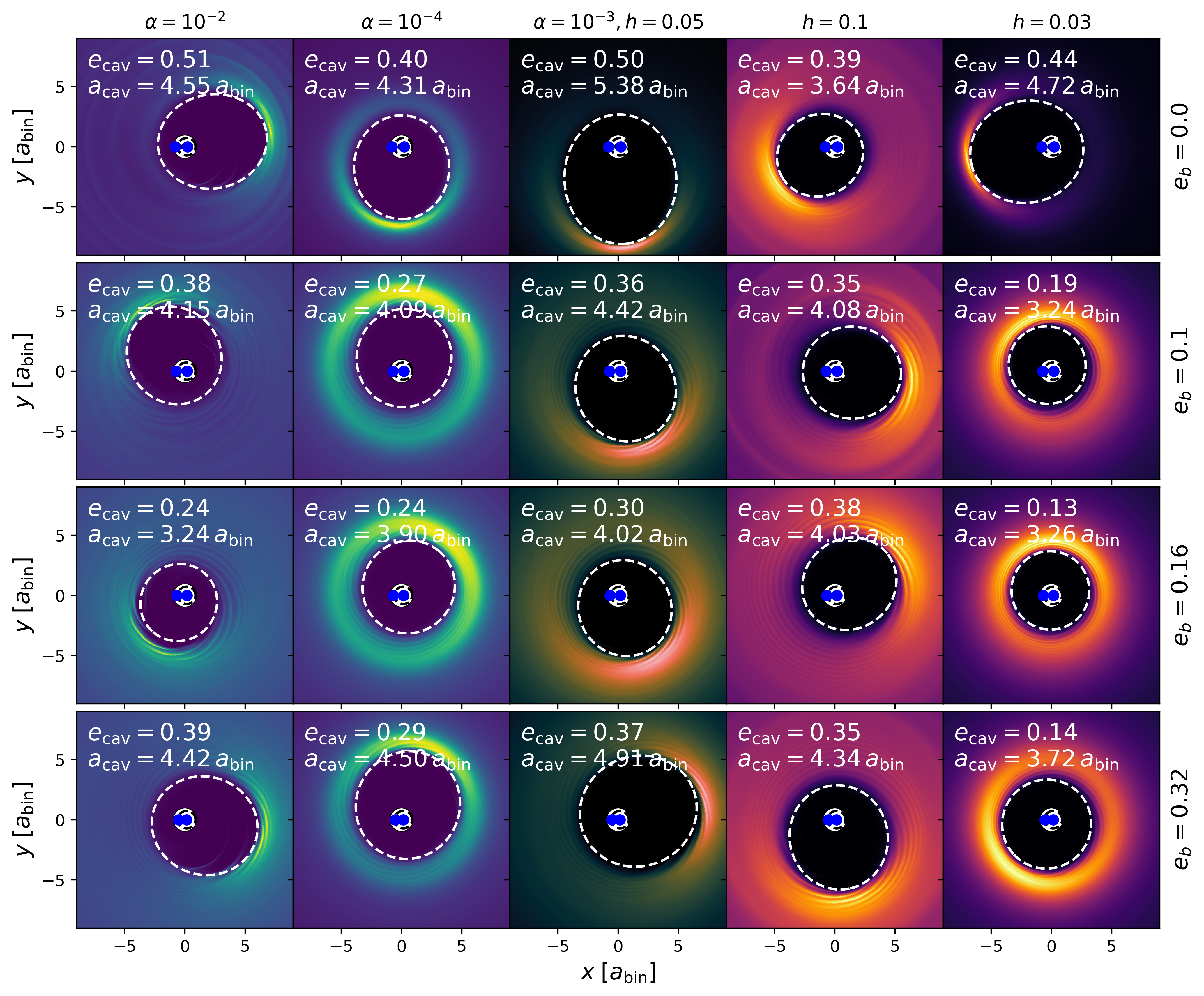}}
		\caption{Map of the gas surface density normalized to maximum surface density for different disc setups varying h (purple maps) and $\alpha$ (green maps) across columns and binary eccentricity across rows. The fiducial setup is shown in the middle column.}
		\label{fig:2d_map}
	\end{figure*}
	
	\begin{figure*}
		\centering
		\resizebox{\hsize}{!}{\includegraphics{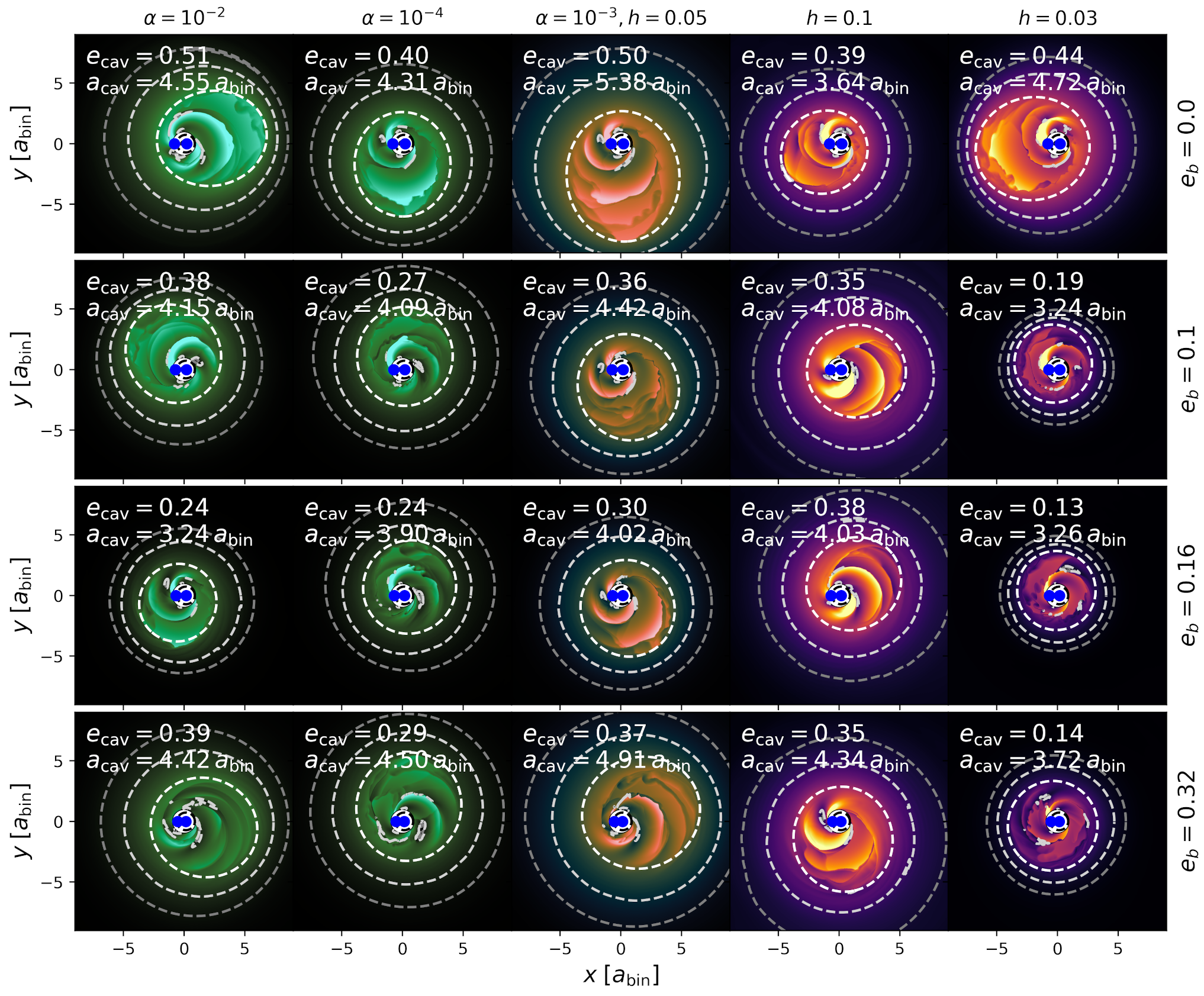}}
		\caption{Map of the dynamic eccentricity for different disc setups varying h (purple) and $\alpha$ (green) across columns and binary eccentricity across rows. The fiducial setup is shown in the middle column. The colour scale ranges from 0 to 1 linearly. The white dashed line marks the cavity, the light grey marks the orbit with half the cavity eccenrticity and the grey line marks the orbit with 1/4 the cavity eccentricity.}
		\label{fig:2d_map_ecc}
	\end{figure*}

\subsection{Effects of disc viscosity}
 
The viscosity in our model is dependent on two variable components: the aspect ratio $h$ and the Shakura-Sunyaev parameter $\alpha$. 
Using this relation has one additional convenient effect of separating $\alpha$ which contributes solely to the viscosity and pressure scale height contributing to both, viscosity and pressure. In addition, the scale height of a disc inferred from observation could be used as a test case to understand if the viscous $\alpha$ model is a good match for the turbulent viscosities in such disc if the binary eccentricity is well known.
 
The cases for the minimum and maximum values of aspect ratio and viscous $\alpha$ are displayed in Fig.~\ref{fig:2d_map} already shown that there is no simple linear relation between the disc shape and either parameter. 
For example, in the case of $\eb=0$ the highest and the lowest $\alpha$ lead to a similar cavity shape, however, the fiducial model inbetween the two creates much larger cavities.
Therefore, we will investigate a range of both values to map the effects of $\alpha$ and $h$ on the cavity shape.
 
    Fig.~\ref{fig:2d_map_ecc} shows the dynamic eccentricity of the gas on its orbit based on Eq.~\ref{eq:ecc}. It shows how far the eccentric modes can perturb into the disc. While this range seems to change little for different $\alpha$-values, reducing the scale height at constant $\alpha$ strongly reduces the reach of the eccentricity into the disc.
	
	\subsubsection{Viscous $\alpha$}\label{sec:alpha}
	
We explored viscous $\alpha$-parameters of $10^{-2}$, $5\times10^{-3}$, $10^{-3}$, $5\times 10^{-4}$, $2.5 \times10^{-4}$ and $10^{-4}$.
From high viscosity to low viscosity the gas density in the disc becomes more concentrated near the inner edge of the disc and the inner disc beyond the cavity circularizes closer to the binary, as shown in Fig.~\ref{fig:2d_map}. For values as low as $\alpha=10^{-4}$ a ring-like over-density forms beyond the cavity. In all cases, the peak density appears at pericentre via the so-called traffic jam effect due to the slower Keplerian velocity at this position.
	
    \begin{figure}
		\resizebox{\hsize}{!}{\includegraphics{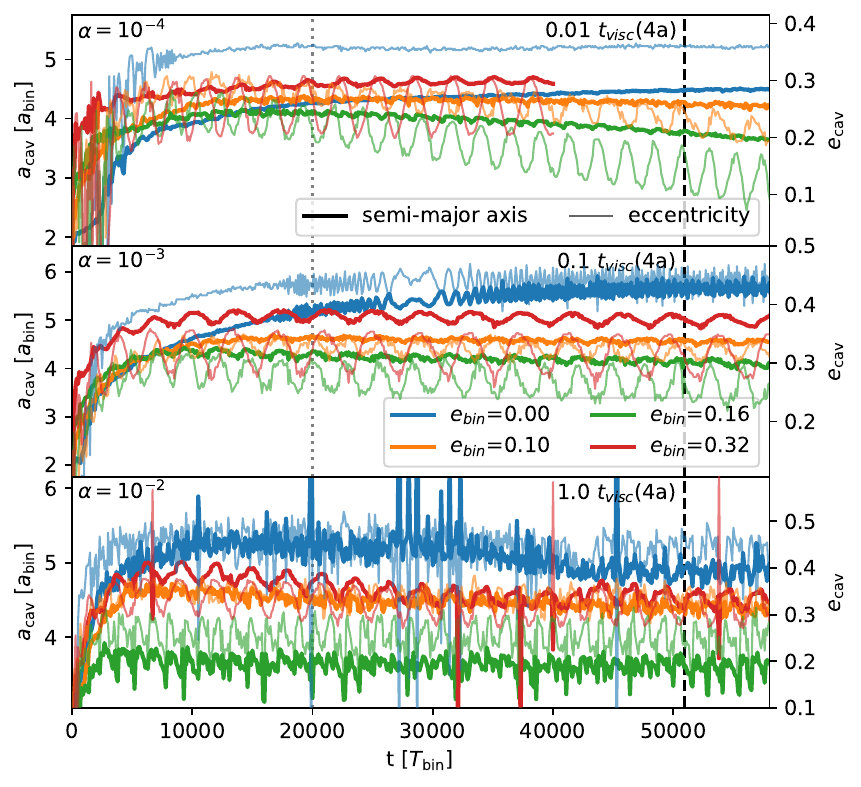}}
		\caption{Cavity evolution of the discs varying $\alpha$ in Fig.~\ref{fig:2d_map}. The thick, saturated line is the semi-major axis of the cavity and the thin, pale line marks the cavity eccentricity. The grey dotted line indicate the point beyond which the data was included in the time average. The dashed black line marks a relative viscous time scale at $4~\ab$ noted in the figure.}
		\label{fig:a_evo}
	\end{figure}
	
	Fig.~\ref{fig:a_evo} shows the evolution of models with different $\alpha$ in Fig.~\ref{fig:2d_map}. All simulations see a steep rise in eccentricity and semi-major axis during the initial $10\,000~\Tb$.
	During this initial raise some systems overshoot and need some time to relax back to the steady state, in particular, the $\alpha=10^{-2}$ cases show this behaviour. The excess gas initially pushed out from the cavity location transports more momentum towards the inner disc region to overexcite it, eventually the steady inflow evens this out. 
    Even though $10\,000~\Tb$ is a slow growth, the time scale for the growth is much faster than viscous times at $4~\ab$ in our simulations, which is reached in the simulation after $5.1 \times 10^4~\Tb$ for our highest $\alpha$.
    Across all viscosities and even compared to studies with more viscous simulations like in \cite{2017Miranda} ($\alpha=h=0.1$) the growth of the disc cavity happens on a comparable timescale of a few thousands of orbits. \cite{2017Miranda} found that the excitation of eccentricity propagates into the disc on even shorter timescales.
    This indicates that other dynamic effects like the for example gravitational interaction and wakes may dominate the behaviour of the cavity and inner disc.
    
	For most simulations, this converges to a steady oscillation by $\sim 20\,000~\Tb$. The strongest evolution beyond this point happens for systems with either low values of $\alpha$ and circular binary orbits, where the semi-major axis of the cavity slowly increases or for binary eccentricities near the bifurcation point of $\eb\sim0.16$, where the disc slowly starts to circularize again. 
    To further explore if the disc renters a steady state accretion disc behaviour, we discuss the mass accretion after the initial growth in the Appendix \ref{sec:app}.
    The oscillation in the orbit parameters occurs on the precession time scale, as also shown in Fig.~\ref{fig:cav_evo} and discussed in Sec.~\ref{sec:short_term}.
	With increasing binary eccentricity, the closest approach between the secondary and cavity edge varies more strongly during the precession of the disc around the binary. Thereby, the oscillation amplitude of cavity parameters also increases with $\eb$.
	
\begin{figure}
    \centering
    \resizebox{\hsize}{!}{\includegraphics{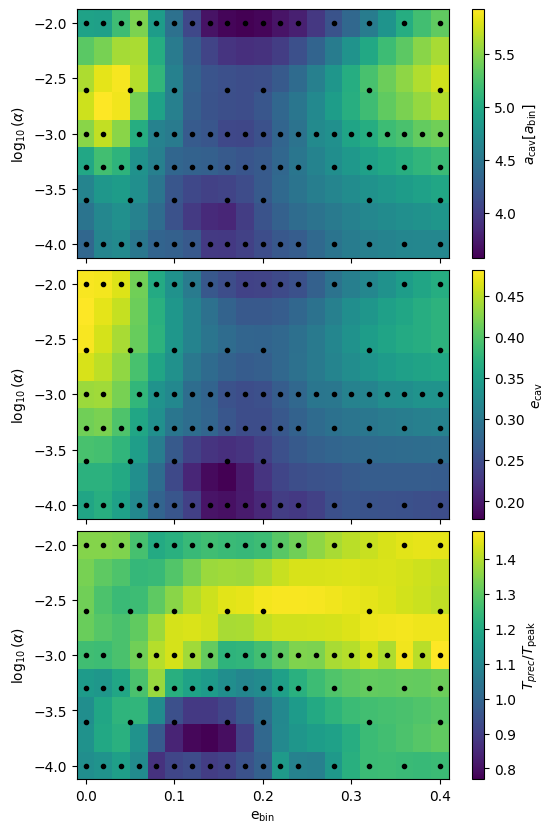}}
    \caption{The cavity size and eccentricity of different binary eccentricities for different viscosities and a scale height of $h=0.05$. Black dots mark simulation setups. Data was linearly interpolated between simulations.}
    \label{fig:alpha_a}
\end{figure}

	The time-averaged effects of the changing viscosity are shown in the parameter map in Fig.~\ref{fig:alpha_a}. 
	The figure shows that the cavity shape is not a linear function of either $\eb$ or $\alpha$ with both parameters showing turning points in the range considered. Furthermore, for $\alpha = 10^{-2}$ cavity sizes show a larger range (from 5.5 to 3.6) than for $\alpha = 10^{-4}$ (from 4.7 to 4).  For constant $\eb$ the largest, most eccentric cavities can be found around an $\alpha$-value of $10^{-2.5}$ and the least eccentric cavities are located between $\eb\sim 0.1 - 0.2$.
	The $\alpha$-parameter is directly linked to the in-drift of gas.
An analytical way to describe the mechanism that would explain the full non-monotonic behaviour has not been presented yet. While the cavities and minimum cavity sizes can be predicted by various ideas like the interplay between viscous and gravitational, resonant torques \citep{2013Dorazio} or the orbital dynamics near the moving binary potential \cite{2023Mahesh} the maximal excitation into the disc depends on hydro- and orbital dynamics in a complex way. The shape difference when altering viscosity in two different ways in Fig.~\ref{fig:2d_map} between varying $\alpha$ or aspect ratio shows that in addition to viscous flow, wave propagation plays a part in shaping the discs.
The maximum turning point for the cavity properties in both parameters (Fig.~\ref{fig:alpha_a} \& \ref{fig:h_a}) separates along a single $\alpha-value$ or aspect ratio, which indicates that the excitation behavior is connected to the viscosity in a non-linear way.
For the lowest $\alpha$, the cavity of discs with $\eb\sim0.16$ can further shrink, which will deepen the bifucation trends across binary eccentricity and viscosity.
For smaller viscosities the precession period is in better agreement with the precession period of test particles located at the peak of the surface density; this is likely a consequence of the ridgid precession being dominated by the narrow inner ring present in these discs (Fig ~\ref{fig:2d_map}).
However, changes are small. For low viscosities and $\eb\approx0.18$ the precession rate reaches its minimum.  
	
	\subsubsection{Aspect ratio and Temperature}\label{sec:h}
	
The temperature is directly derivable from the aspect ratio. For the most massive and expanded disc like GG Tau the size of the system be irradiation-dominated with larger scale heights -- in the case of GG Tau $h\sim0.1$ \citep{1999Guilloteau}.
In general, binary systems typically receive less luminosity per central mass because the same mass is split into two less luminous stars. 
Therefore, colder systems with a correspondingly low aspect ratio are reasonable at smaller separations like in CS Cha \citep{2022Nico}. 
In previous studies on viscously-heated circumbinary discs \citep{2019Kley, 2022Sudarshan}, we have shown that viscous heating produces systems with an aspect ratio between 0.04 and 0.05 for the inner part of the disc at $\sim3.5~\ab$ for close binaries ($\ab \sim 0.2~\mathrm{AU}$) comparable to the observed planet-hosting Kepler binary systems. 
In the following, we have varied the aspect ratio from 0.03 to 0.1. For $\ab=0.2~\mathrm{AU}$ around a solar-mass binary, this corresponds to a temperature of about $\sim100\,\mathrm{K}$ to $\sim800\,\mathrm{K}$ at $3\,\ab$.
	
	\begin{figure}
		\resizebox{\hsize}{!}{\includegraphics{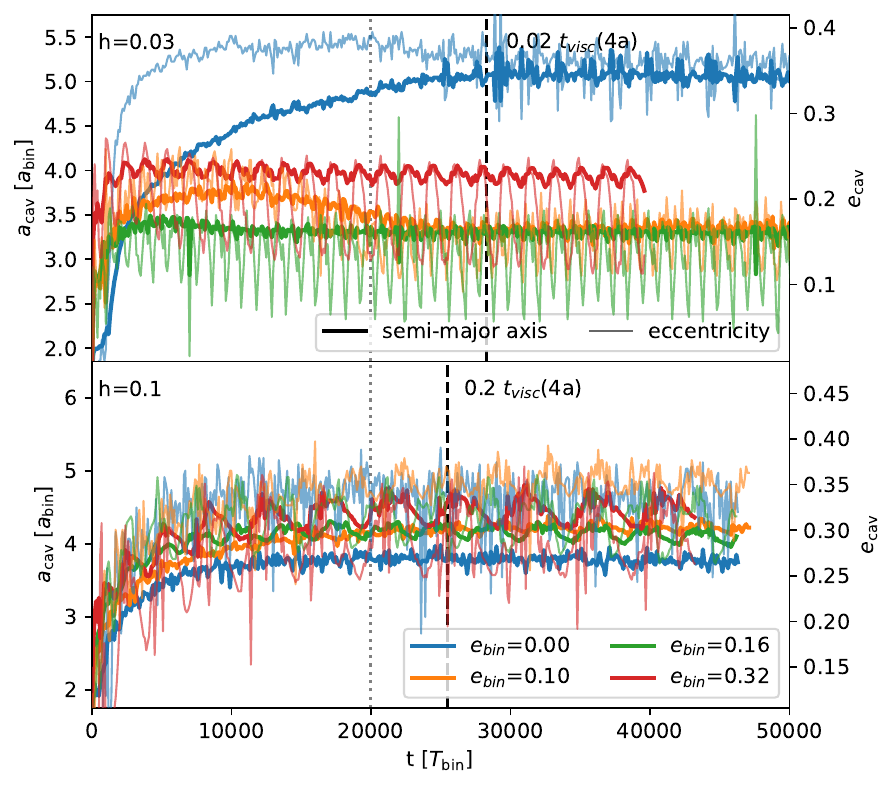}}
		\caption{Cavity evolution of the discs varying $h$ in Fig.~\ref{fig:2d_map}. The thick, saturated line is the semi-major axis of the cavity and the thin, pale line marks the cavity eccentricity. The grey dotted line indicate the point beyond which the data was included in the time average. The dashed black line marks a relative viscous time scale at $4~\ab$ noted in the figure.}
		\label{fig:h_evo}
	\end{figure}
	
The time evolution of the discs with different aspect ratio and $\alpha = 10^{-3}$ (see Fig.~\ref{fig:2d_map}) is shown in Fig.~\ref{fig:h_evo}.
For low aspect ratios, the simulations again show an overshooting of eccentricity and cavity size and, similar to the low $\alpha$ case, the disc relaxes into a steady state more slowly.
After $\sim30\,000~\Tb$, the discs reach a converged state for all aspect ratios. 
The cavity size and eccentricity oscillates more strongly for large eccentricities and aspect ratios, and also show complex behaviours.
Disc-binary interaction through the cavity material depends on the aspect ratio through viscous accretion. So at higher aspect ratio and pressures the stream have more mass to alter the cavity shape.
	
\begin{figure}
		\centering
		\resizebox{0.975\hsize}{!}{\includegraphics{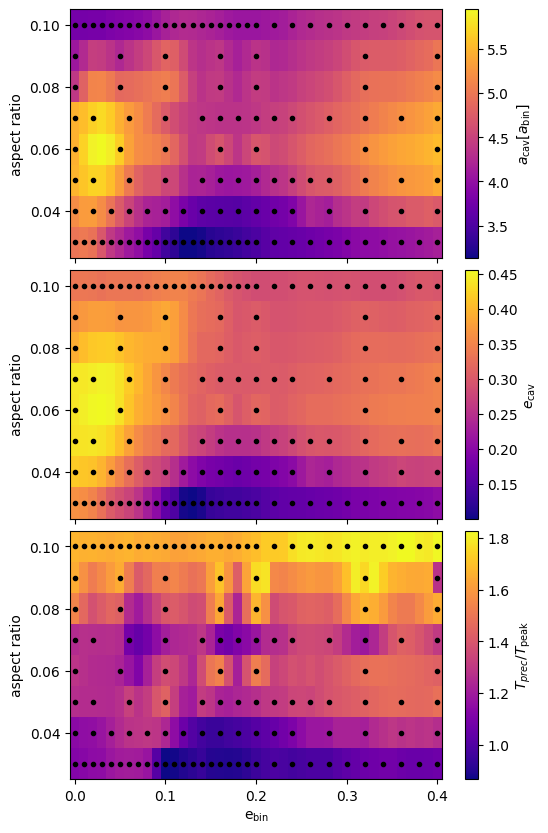}}
		\caption{The cavity size and eccentricity and disc precession scale to the binary's orbit location of different binary eccentricities for different aspect ratios and a viscous $\alpha=10^{-3}$. Black dots mark simulation setups. Data was linearly interpolated between simulations.}
		\label{fig:h_a}
	\end{figure}

We performed a series of simulations for aspect ratio between 0.03 and 0.10, and binary eccentricities between 0.0 and 0.4. The number of different binary eccentricities differ but include at least 0.0, 0.05, 0.1, 0.16, 0.2, 0.32 and 0.4. All simulations were run for at least $30\,000$ orbits.
We averaged the disc properties (see Fig.~\ref{fig:h_evo}) from $20\,000~\Tb$ onwards.
The influence of the aspect ratio is shown in Figure \ref{fig:h_a}.

One consistent trend with increasing aspect ratio is the decreasing variation in cavity sizes over binary eccentricity. While for h=0.03 the size range is a factor of about 2 $a_{bin}$, for h=0.1 they only range from 3.8 to 4.6 $a_{bin}$ which is less than a factor of $20\%$. We find a turning point in cavity size at a scale height of 0.06. Here the cavity size is largest and decreases with increasing and decreasing aspect ratio.
As seen in previous studies, the size reaches a local minimum at $\eb\sim 0.15 - 0.2$ independent of scale height. For the highest aspect ratios, similarly small cavities around circular binaries as well.

The cavity eccentricity increase with h until $h \sim 0.06$, where it plateaus, except for nearly circular binaries. 
The high aspect ratios lead to little variation in the cavity eccentricity of $\sim0.32\pm0.02$a for $h=0.1$. 
The local minimum around at $\eb\sim 0.15 - 0.2$ also occurs in the eccentricity of the disc. This means that circumbinary discs around $\eb\sim 0.15 - 0.2$ are smaller and more circular than other circular or more eccentric binaries independent of scale height.

The peak normalized precession rate increases with a higher aspect ratio, which was already to be expected looking at the extent of eccentricity in the discs in Fig.~\ref{fig:2d_map_ecc}. For higher aspect ratio/pressure, the region of rigid precession covers more of the disc, while it only affects a narrow ring in the case of a low aspect ratio. 
Hence, the precession time is affected by and reflects an orbit further out the disc for higher aspect ratios.
Changes due to varying the aspect ratio are more significant than changes to the $\alpha$ viscosity.  
	
As the cavity parameters are less sensitive to system parameters at larger aspect ratios, the differences due to the binary eccentricity expected in hot, young, discs or discs around wide binaries that are flared by irradiation are expected to be small. But these variations become crucial for understanding cooler, evolved discs during or towards the end of planet formation and migration or discs around close binaries in which the circumbinary Kepler planets \citep[e.g.][]{2021TIC} formed.
	
\subsubsection{Thick discs}

    Especially for very large discs ($\ab > 20~\mathrm{au}$) the flaring due to the irradiation of the central stars leads to high aspect ratios, for example, $h=0.11$ for GG Tau \citep{1999Guilloteau} or $h=0.1$ for HD142527\citep{2021HD142527}. At the same time, the irradiation dominates the temperature profile. This leads to conditions with fast cooling that are comparable to locally-isothermal conditions studied here. 
    Such systems are easily observable too, which makes them an interesting test case for understanding the circumbinary disc at this regime. Therefore we have conducted runs with $\alpha = [10^{-4},10^{-3},10^{-2}]$
    and varying $\eb$ for these discs.

\begin{figure}
		\centering
		\resizebox{\hsize}{!}{\includegraphics{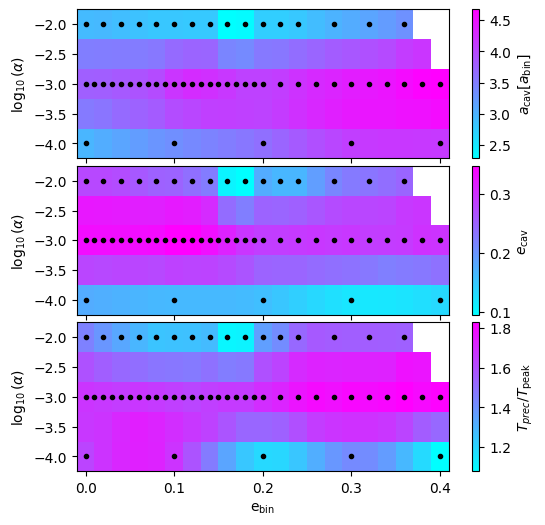}}
		\caption{The cavity size and eccentricity and disc precession scale to the peak's orbit location of different binary eccentricities for aspect ratio = 0.1 and visous $\alpha=[10^{-4}, 10^{-3}, 10^{-2}]$. Black dots mark simulation setups. Data was linearly interpolated between simulations.}
		\label{fig:a_ex}
\end{figure}

The shapes of these discs in Fig.~\ref{fig:a_ex} show that overall the high pressure leads to a reduction in the range of the eccentricities to only between $e_\mathrm{cav} = [0.1, 0.33]$ whereas near for h=0.05 the cavity eccentricity reached up to 0.45. Compared to the same $\alpha$-values with $h=0.05$ the cavity size reduces from $a_\mathrm{cav}(h=0.05) = [3.6, 6.0]$ to $a_\mathrm{max}(h=0.1) = [2.5, 4.5]$. The disc's precession rate is longer than the n-body orbit at the peak location through all simulations, indicating a rigid precession deeper into the disc. 
The highest viscosity, $\alpha=10^{-2}$, reduces the size of the cavity significantly to $a_\mathrm{peak}\approx3.5~\ab$ with a slight increase toward higher $\eb$. 
The disc accretes on short time scales as the overall dynamic viscosity is high ($\nu\sim10^{-4}~\ab^2\Omega_\mathrm{bin}$). As a result, these models cover a similar parameters space to the models by \cite{2022Dittmann}. 
Such a high viscosity, however, would require a very significant level of turbulence in the system that most mechanisms operating in protoplanetary discs cannot achieve (except maybe close to the stars, where $h < 0.3$).

A viscosity of $\alpha = 10^{-3}$ leads to significantly larger inner cavities ($\geq ~3.6\ab$) which are more eccentric throughout and have an increasing eccentricity for decreasing $\eb<0.16$ below the bifucation point.
    
For even lower viscosity , $\alpha = 10^{-4}$, the cavity shrinks again as in Fig.~\ref{fig:alpha_a}. Interestingly, for such inviscid thick discs the bifucation behaviour breaks and cavity sizes increase steadily with binary eccentricity while the disc eccentricity remains on the same level at $\sim 0.15$. Such a change for the circular binaries at high pressures might be related to the same conditions as noted for high pressures due to long cooling times in \cite{2022Sudarshan}.

\section{The curious case of HD 142527}\label{sec:exa}

We know of some examples of large, asymmetric disc cavities around wide binaries like GG-Tau \citep[which we simulated in][]{2020Keppler} or HD 142527 \citep{2006Fukagawa,2011Verhoeff,2021HD142527}.

To reproduce such large and eccentric cavities in general,
simulations in regimes of thick disc with a viscous $\alpha=10^{-3}$ is preferable. It creates the largest cavity semimajor axis, and this semimajor axis increases further with binary eccentricity. 
The cavity eccentricity reaches a level of $\geq0.2$ and the binary-driven precession reaches deep into the disc.

\begin{figure}
		\centering
		\resizebox{\hsize}{!}{\includegraphics{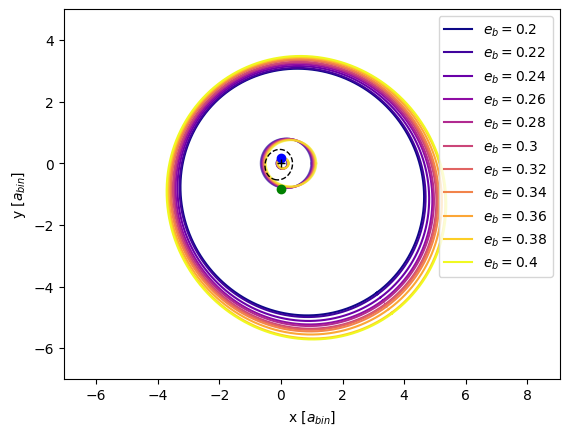}}
		\caption{The position of the cavity edge (density drops to 10\% of the peak density) and the binary orbit of the binary for $h=0.1$, $\alpha=10^{-3}$ and $\eb=[0.2,0.4]$.
        The black dashed line indicates the best-fit orbit of HD 142527 by \protect\cite{2024Nowak} relative to the cavity size.}
		\label{fig:hd_level}
	\end{figure}

HD 1425227 is a binary system of an early Herbig Star and a red dwarf surrounded by a disc with an inner cavity of 10 times the current distance of the binary stars \citep{2019Claudi,2024Nowak}.
The disc around HD142527 has been mysterious due to its immense cavity of 100-140~au and its clearly asymetric structure. 
This led in the past to speculations about possible orbit configurations e.g. in \cite{2018Price} invoking highly eccentric binary orbits with $\eb>0.6$ to achieve a cavity size of $a_\mathrm{cav}\sim3~\ab$ to explain the first estimates of the stellar distances known at that time.
Observations of the disc point to a scale height of $h\approx0.1$ \citep{2021HD142527}. We therefore look at the thick disc case in Fig.~\ref{fig:a_ex}.
Even with lower $\eb$ our discs reach larger sizes ($a_\mathrm{cav}\geq4~\ab$) after $>30~$k$\Tb$ evolution time to reach the disc full cavity size.
Fig.~\ref{fig:hd_level} shows how the density maximum and the binary orbits would shift due to the change in eccentricity above $\eb>0.2$. 
Increasing eccentricities would further widen the cavity, such that it is reasonable to see a gap with sizes of $\geq10$ times the pericentre separation of binary stars.

These scales would have agreed well with previous orbit best-fits \citep{2019Claudi} that suggested a pericentre distance $\geq18$~au such that the simulations could reach a cavity size of 140~au.
However, the latest work by \cite{2024Nowak} constrains the binary orbit to $\ab=10.8\pm0.22~$au with $\eb=0.47\pm0.01$.

Therefore, our simulations are also in disagreement with the new GRAVITY measurements of a semi-major axis of 10.8~au for which none of the simulations produce cavity sizes above $\sim50~au$. Although this can vary by $\sim10\%$ as the disc precesses, it is insufficient to explain the size of a cavity $\geq 10\times$ the separation of the binary.
The approximate scale of the real orbit relative to the cavity size is marked in Fig.~\ref{fig:hd_level} and it is about half of the size of the previous estimate.
Within the \emph{entire} parameter space considered here, such a large separation of the cavity is impossible to reproduce in our simulations. The largest possible size would be reached with a lower aspect ratio of $h=0.06$ with a cavity size of $\leq70$~au around a binary with $\eb \leq 0.4$. As a result, additional effects are likely needed to explain the cavity size, with effects such as winds, more complex internal heating and cooling of the disc or other objects within the disc all possibly changing the cavity size.
While the eccentric disc suggests an impact of the binary interaction of the disc it is not the sole cause of the size of the inner cavity.

\section{Artifical profiles}\label{sec:art}
Using the parameter map that we created in section \ref{sec:result_II} we can make predictions about the disc shape to inform how we initialize circumbinary disc simulations or how we construct a simple circumbinary disc density profile taking the asymmetry into account. This can help to speed up simulation convergence and to create radiative transfer comparisons of observed systems to better understand underlining system parameters.

To achieve such a comparable profile we create density maps following a few simple assumptions. First, all material orbits on an eccentric orbit around the center of mass that marks the focal point c and have the same argument of pericentre of 0. Second, the eccentricity steadily reduces outwards. Third, the eccentricity $e_\mathrm{cav}$ and semimajor axis $a_\mathrm{cav}$ of the cavity edge are known. 
With this it follows for any position (x,y) in the disc that their eccentricity $e$ and semimajor axis $a$ can be calculated with

\begin{align}
a(x,y) =& \Bigg( \frac{(x-c)^2+y^2+c^2}{2} \\ \nonumber
		&+ \sqrt{\frac{((x-c)^2+y^2+c^2)^2}{4} - c^2(x-c)^2} \Bigg) ^{1/2} \\
&\mathrm{with} \quad c = a_\mathrm{cav}e_\mathrm{cav} = e(x,y)a(x,y).
\end{align}

Knowing a local semimajor axis allows to calculate the elliptic Keplerian speed

\begin{equation}
v^2(x,y) = M_\mathrm{bin}G \left(\frac{2}{\sqrt{x^2+y^2}}-\frac{1}{a(x,y)}\right) \,.
\end{equation}

Assuming that the density maximum in the apocenter of the disc is caused by a traffic jam of the gas the density structure around the disc is rescaled to match the inverse of the velocity and the initial cavity is included,

\begin{equation}
\Sigma_\mathrm{el}(x,y) = \Sigma_\mathrm{ref} \left(\frac{a}{r_\mathrm{ref}}\right)^{-1.5} \left[1 + \exp(-\frac{a-a_\mathrm{cav}}{0.1 a_\mathrm{cav}})\right]^{-1} \frac{v_K}{v}.
\end{equation}

This profile will nicely recreate an eccentric inner density distribution. However, the eccentricity only converges to 0 at infinity while in viscous discs the dynamics on the outer disc will not be dominated by the eccentric mode of the binary. To account for such an eccentricity cut off, outer disc can be transitioned into a symmetric disc at outer radius $r_\mathrm{sym}$ using 
\begin{align}
\Sigma_\mathrm{art} =& (1+\exp(k))^{-1}\Sigma_\mathrm{el}+ (1+\exp(-k))^{-1} \Sigma\\ \nonumber
&\mathrm{with} \quad k =  \frac{r - r_\mathrm{sym}}{0.1 r_\mathrm{sym}} \, .
\end{align}

\begin{figure}
    \centering
    \resizebox{\hsize}{!}{\includegraphics{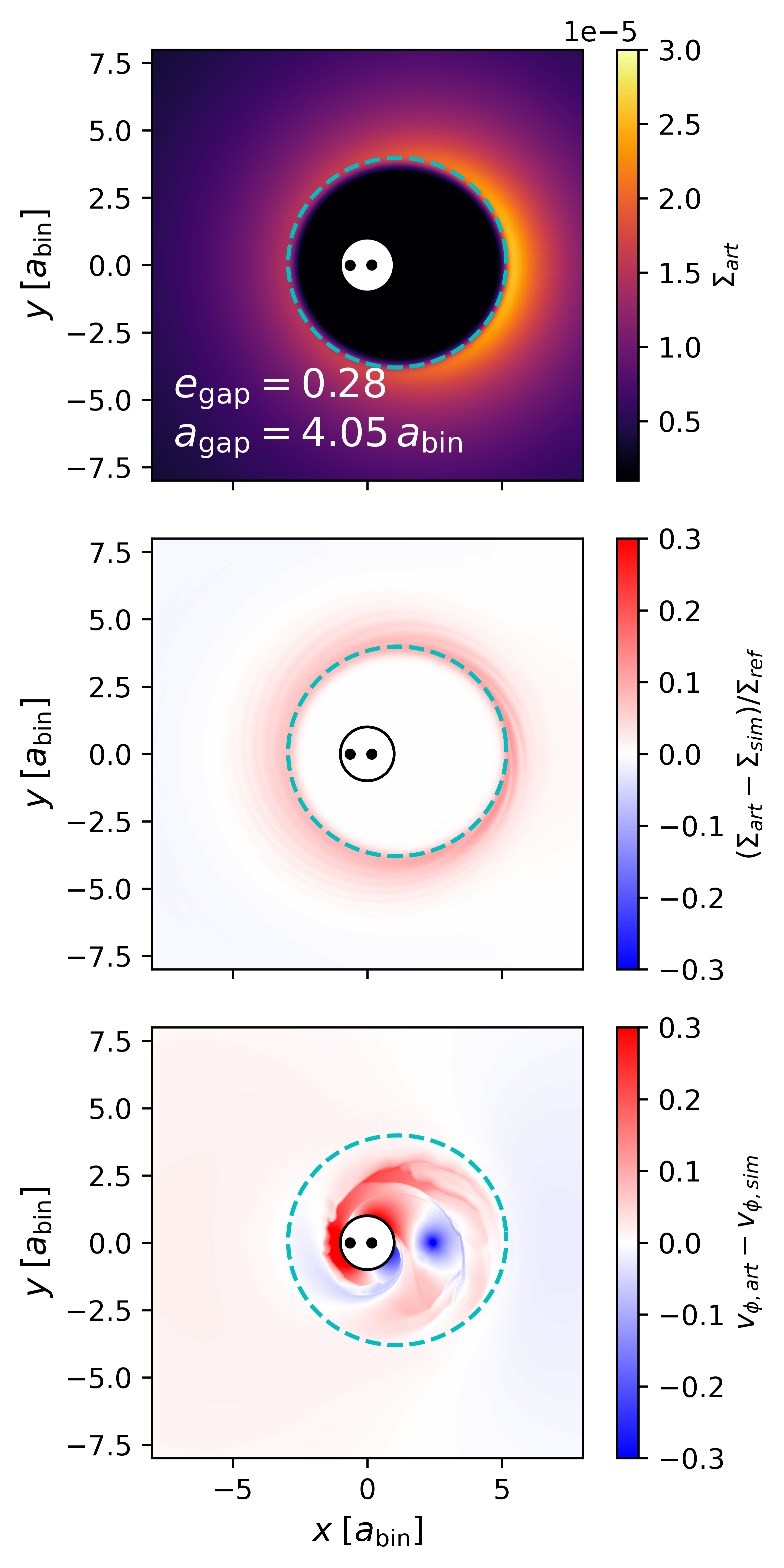}}
    \caption{Comparison of the artificial and simulated disc profile.}
    \label{fig:art_ini}
\end{figure}

As a first-order approach for a set of initial conditions, $v_\phi$ is set to $v$ and $v_\mathrm{r}$ to 0. For a better solution, both velocity components can be derived using the conservation of angular momentum. For a comparison we used $r_\mathrm{sym} = 15\,\ab$. In Fig.~\ref{fig:art_ini}, this description is compared to the standard case after $80.5\, kT_\mathrm{bin}$. In the density, there is a $\sim 10\%$ deviation at the edge position but the general shape of the main part of the outer disc is well reproducible. In the inner disc there are some deviations inside the cavity where the motion of the binary perturbs the gas, for the main disc the eccentric velocities are a good initial approximation of the disc.
When altering the aspect ratio $r_\mathrm{sym}$ should be adjusted to match the size of the rigidly bound disc.

\begin{figure}
    \centering
    \resizebox{\hsize}{!}{\includegraphics{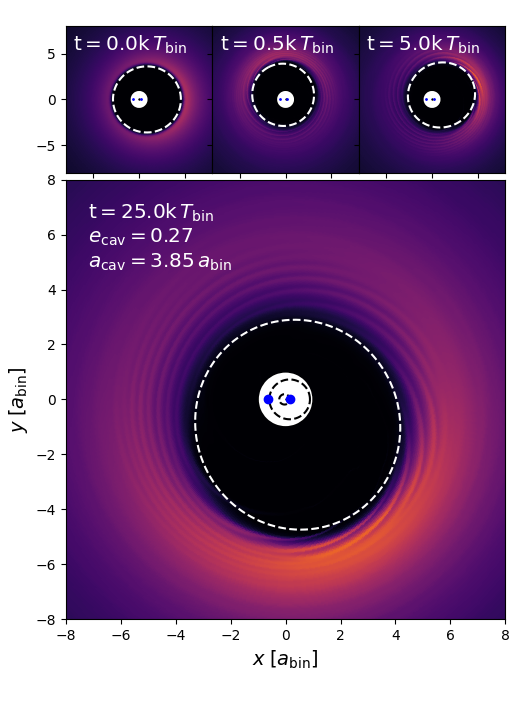}}
    \caption{Surface evolution started from a initial elliptic profile.}
    \label{fig:art_evo}
\end{figure}

When we tested this gas distribution and velocity map as initial conditions, we could reduce the time needed to reach convergence by a factor of 2. This means it still requires a long simulation time of the order of $10^4~\Tb$ depending on the disc parameters. 
The motion of the binary interacts with the disc by wakes of viscous gas through the inner cavity leading to a slow build-up of the eccentricity.
In Figure~\ref{fig:art_evo} the snapshot shows the evolution of the artificial profile towards the final disc structure. Initially, the momentum missing due to the absence of spiral wakes leads to a reduction in the eccentricity relative to the input value in the first few thousand orbits. While the analytic model, unfortunately, may not be the ideal solution to study the dynamics of circumbinary discs quickly, it does allow quick simulations and simple models to compare specific disc shapes to observational data.

\section{Discussion}

The models in this work are strongly simplified to reduce the parametrization to a few simple parameters and scales related to binary properties rather than direct physical scales. While this approach makes the results relevant for a wide range of circumbinary discs, it neglects effects that can influence the viscosity and thermodynamic structure of real circumbinary discs, such as radially flaring disc scale heights  \citep{2020Ragusa} due to possible irradiation heating, radiative cooling \citep{2019Kley} or the $\beta$-prescription of cooling scaled to the orbital time \citep{2022Sudarshan, 2023Wang}, magneto-hydrodynamics \citep{2012Shi}, self-gravity \citep{2017Mutter}, winds, and the effects of dust \citep{2020Pierens, 2022Coleman}. 

As we used a locally-isothermal model for the disc we did not include irradiation or viscous heating. In previous studies \citep{2019Kley} we found that viscous heating is the dominant heat source in the conditions relevant to the known circumbinary planet-hosting systems. Those results produced temperature profiles that resemble the flat aspect ratio we used in this study. Further, \cite{2019Kley} and \cite{2022Sudarshan} have shown that viscous heating and radiative cooling create inner cavity structures comparable to locally-isothermal models. The reason for this similarity is the short, nearly isothermal, cooling times in the optically thin cavity. However, when cooling times inside the cavity are long or near the orbital time scale this can be suppressed and the disc cavity becomes nearly circular and limited to the unstable region \citep[see][]{2022Sudarshan,2023Wang}.

For systems in which irradiation becomes a dominant factor (e.g. $\ab\sim 5~$au), a flared disc height would be a better approximation \citep[see e.g.][]{2020Ragusa}.
The next-order approach is to include the heating and cooling due to radiation in the model.
Depending on the conditions, the discs can have or transition through cooling times that are of order of the orbital time scale. Under these conditions, cooling can strongly damp waves \citep{2020Miranda, 2023Alex} and lead to a reduced eccentricity growth.
This requires binary separations of several au for sun-like stars and a relatively small range of optical depths, which may make these effects important in only a small parameter space. The precise conditions in which discs are affected and how this affects the shape of the circumbinary disc will be the subject of a future study.

At the minimum for cavity sizes ($\eb\approx0.16$) other studies like \cite{2021Dorazio} and \cite{2023siwekI} find complete recircularization of the discs for higher viscosities of $\nu=10^{-3}~\ab\Omega$ and $\alpha=0.1$. In the locally isothermal, low-viscous PPD environment this is not the case, even though we can not exclude it for the lowest $\alpha$. To reach a circularizing condition the exciting wakes which the binary creates need to be overpowered by the advective disc flow.
Recircularization can also be a transient, meta-stable behaviour as discussed in \cite{2020Keppler} that only last up to a few thousand binary orbits. In other test, we have see this to occur more easily in systems with more even mass ratios between the binary components at early times in the simulations.

For this large study, only 2D simulations were feasible. However, 3D effects \citep{2012Shi, 2020Pierens, 2021Pierens} can change the results, as the interaction between streams and discs is no longer forcefully aligned in-plane and the turbulence driven by the eccentricity, becomes evident in 3D simulations. 
The long convergence time would make achieving steady-state in 3D simulations extremely expensive and thus was beyond the scope of a parameter study. However, the results presented can be used as the starting point for a 3D model since a good initial starting guess reduces artefacts and shortens the convergence time.

Since this work is focussed on the structure of the disc, we do not discuss the effects of the disc onto the binary, to predict its orbital evolution \citep{2016Rafikov}. This subject is mostly studied in the context supermassive black holes \citep[e.g.][]{2017Miranda,2021Tiede,2021Dittmann}. The binary orbital evolution in the PPD will be the subject of the second upcoming publication.

\section{Conclusion}

With this work, we provide a guide towards unresolved binary stars and invisible turbulent disc viscosity using the interaction of binary stars with their protoplanetary discs. 
To understand the parameter realm of circumbinary disc dynamics, we created $>140$ long-term, locally isothermal 2D simulations.
By mapping the binary eccentricity with the aspect ratio and the scale height an observed disc structure can be translated into these parameters. 
There is no linear trend arising from any of the parameters, but there are turning points to understand regimes of the behaviour:
\begin{itemize} 
	\item Independent of viscosity, the size and eccentricity of the inner cavity is minimal at a binary eccentricity $\eb \sim 0.18$, reproducing the bimodal behaviour of \cite{2017Thun} for the whole range of parameters.
	\item For both viscous $\alpha$ and $h$ there exists values of maximum cavity size and eccentricity, at $\alpha\approx 3\times10^{-3}$ and $h\approx 0.06$. Away from this value the excitation of the inner cavity is either limited by the low accretion or by fast inflow from the outer symmetric disc.
	\item For increasing viscosities, 
    the excitation reaches further into the disc, especially when increasing the aspect ratio. This forces a rigid precession through a large region of the inner disc. 
\end{itemize}

The simple parameters of $a_\mathrm{cav}$ and $e_\mathrm{cav}$ can be used to construct artificial disc profiles, These profiles can be used to create gas and dust distribution for comparison with observations or for initial conditions to reduce the simulation time to reach the final configuration by a factor of a few.

We compared our extensive grid of simulations to the newest orbital constraints from HD 142527 \citep{2024Nowak}. These results confirm the difficulty explaining the cavity size and shape reported by \cite{2024Nowak}, despite the long run times and steady-state disc structures having larger cavity sizes than those previously reported \citep{2018Price}. It is therefore likely impossible to explain the observed disc structure with isothermal, viscous circumbinary discs and additional physics such as MHD, radiative heating and cooling, or additional compansions are needed.

\section*{Acknowledgements}
This work would not have been possible without Willy Kley and there are no good enough words to thank him for his guidance and kindness we remember him by. 
We thank Daniel Thun for his initial contribution and help in using the GPU-Pluto version and James Owen for helpful discussion.
AP and RAB acknowledge support from the Royal Society in the form of a University Research Fellowship and Enhanced Expenses Award.
Parts of this study were funded in part by grant KL 650/26 of the German Research Foundation (DFG).
This work was performed with the support of the High Performance and Cloud Computing Group at the Zentrum f\"ur Datenverarbeitung of the University of T\"ubingen, the state of Baden-W\"urttemberg through bwHPC and the German Research Foundation (DFG) through grant no INST 37/935-1 FUGG, and Cambridge Service for Data Driven Discovery (CSD3), part of which is operated by the University of Cambridge Research Computing on behalf of the STFC DiRAC HPC Facility (www.dirac.ac.uk). The DiRAC component of CSD3 was funded by BEIS capital funding via STFC capital grants ST/P002307/1 and ST/R002452/1 and STFC operations grant ST/R00689X/1. DiRAC is part of the National e-Infrastructure.
RPN acknowledges support for the Leverhulme Trust grant RPG-2018-418 and STFC grants ST/X000931/1 and ST/T000341/1.

\section*{Data Availability}
The data underlying this article will be shared on reasonable request to the corresponding author.

\bibliography{parameter_study}
\bibliographystyle{mnras}

\appendix 
\section{Convergent solution} \label{sec:app}
Even though the runtimes of the simulations are long, only the models with the highest viscosity reach viscous times in the disc, because of the low viscosities acting in protoplanetary environments as already indicated in Fig.~\ref{fig:a_evo}~\&~\ref{fig:h_evo}. Nevertheless, the state of the circumbinary cavity and inner disc is not created through a viscous but gravitational interaction, which does not need to act on a similar time scale. 

\begin{figure}
    \centering
    \resizebox{\hsize}{!}{\includegraphics{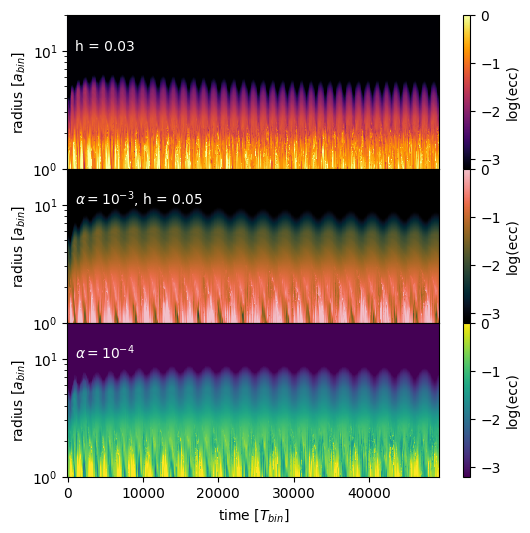}}
    \caption{Density weight, azimuthally averages eccentricity evolution throughout the disc. All models have a $\eb=0.2$. The colors indicate the orbital eccentricity. The top panel shows the lowest aspect ratio $h=0.03$, the mid panel shows the fidicial model $\alpha=10^{-3};~h=0.05$ and the bottom panel shows the lowest $\alpha$-value of $10^{-4}$.}
    \label{fig:ecc_time}
\end{figure}

The disc react to the perturbation cause by the binary on a time scale short than the viscous time deep into the disc. This can be see in the average eccentricity grow in the disc, in Fig.~\ref{fig:ecc_time}. The eccentricity in the disc grows within $>10\,000~\Tb$ and after that continues to vary with the precession time scale of the disc. This is consistent with the models by \cite{2017Miranda} and \cite{2019Munoz}.

The viscous accretion provides merely a steady background condition for the interaction of symmetric steady inflow pushing onto the eccentric cavity and feeding the exciting spiral outflows within the cavity.

\begin{figure}
    \centering
    \resizebox{\hsize}{!}{\includegraphics{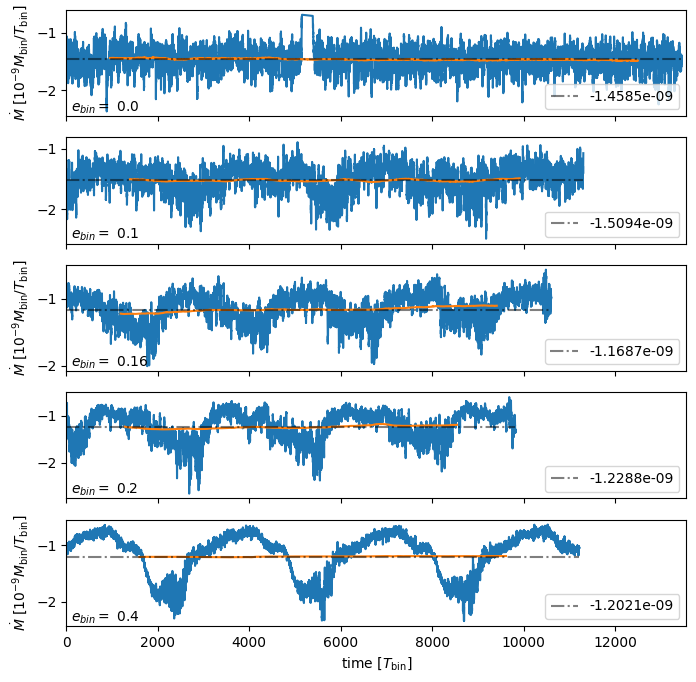}}
    \caption{Mass accretion rate of the converged state of simulations at $\alpha=10^{-4}$. The blue line is the rolling average over $100~\Tb$, the orange line is a rolling average over one precession time and the grey dashed line indicates the average value.}
    \label{fig:mdot_a}
\end{figure}

\begin{figure}
    \centering
    \resizebox{\hsize}{!}{\includegraphics{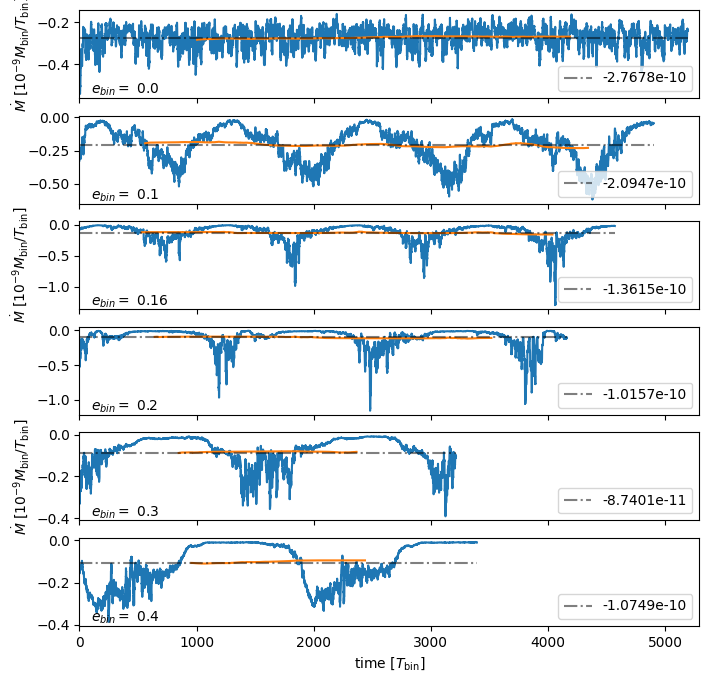}}
    \caption{Same as Fig.~\ref{fig:mdot_a} for $h=0.03$}
    \label{fig:mdot_h}
\end{figure}

To test, if in the data used for the averages, the viscous disc indeed provides such a steady environment we looked at the total mass accretion through the inner boundary for the two extremely low viscosity models of $\alpha=10^{-4}$ in Fig.~\ref{fig:mdot_a} and $h=0.03$ in Fig.~\ref{fig:mdot_h}. The data of these after figures taken from continued simulations with a higher time resolution after the cavity reach a convergent state. 
The accretion in a binary depends on the alignment between the binary orbit and the cavity which will be discussed in much more detail in the second part publication. Thereby, the accretion shows a periodic behaviour if not corrected.
If the accretion is averaged over the precession the lines become nearly constant showing steady accretion at least on the order of $\sim10^4\Tb$.
The initial redistribution of gas and chosen surface density slope do not allow to predict the true value of $\Sigma_0$ for the simulations. However, throughout binary eccentricities, the mass accretion has similar values hinting at a consistent accretion behaviour of the models with the same viscous background. If $\Sigma_0$ is recalculated from the accretion rates if would within ranges of $5\times[10^{-4},10^{-5}] ~M_\mathrm{bin}/\ab^2$, this does not deviate strongly from the initial condition set for the disc. 

The most challenging set of simulations and the only one with an increase of mass accretion compared to the initial setup is $\alpha=10^{-4}$. In this case, the access material inside the cavity set by the initial conditions feeds back material to the disc from the very low density steady cavity region in such a low viscosity case. This creates an overdensed structure in the inner disc that can only slowly disappear through advection of material through the cavity. Nevertheless, the behaviour of the cavity is steady independent on the structure of the outer disc as was already shown in \cite{2020Munoz}.

\bsp	
\label{lastpage}
\end{document}